\documentclass[prl,nofootinbib,twocolumn,preprintnumbers]{revtex4}

\usepackage{color}
\usepackage{graphicx}
\usepackage{amsmath}

\preprint{KUNS-2633}

\begin{document}

\title{Constraining $f(R)$ Gravity with Planck Sunyaev-Zel'dovich Clusters}

\author{Simone Peirone$^1$}
\author{Marco Raveri$^{2,3,4,5}$}
\author{Matteo Viel$^{3,4}$}
\author{Stefano Borgani$^{1,3,4}$}
\author{Stefano Ansoldi$^{6,3,7}$}

\smallskip
\affiliation{$^1$ Astronomy Unit, Department of Physics, University of Trieste, Via Tiepolo 11, I-34131, Trieste, Italy\\
$^{2}$ SISSA - International School for Advanced Studies, Via Bonomea 265, I-34136, Trieste, Italy\\
$^{3}$ INFN, Sezione di Trieste, Via Valerio 2, I-34127 Trieste, Italy\\
$^{4}$ INAF-Osservatorio Astronomico di Trieste, Via G.B. Tiepolo 11, I-34131 Trieste, Italy\\
$^{5}$ Institute Lorentz, Leiden University, PO Box 9506, Leiden 2300 RA, The Netherlands\\
$^{6}$ Department of Physics, Kyoto University, Kyoto 606-8502, Japan\\
$^{7}$ Department of Mathematics, Computer Science and Physics, Udine University, Italy}

\begin{abstract}
  Clusters of galaxies have the potential of providing powerful
  constraints on possible deviations from General Relativity. We use the catalogue of
  Sunyaev-Zel'dovich sources detected by \emph{Planck} and
  consider a correction to the halo mass function for a $f(R)$ class
  of modified gravity models, which has been recently found to
  reproduce well results from $N$-body simulations, to place constraints
  on the scalaron field amplitude at the present time, $f_{R}^0$. We
  find that applying this correction to different calibrations of
  the halo mass function produces upper bounds on $f_{R}^0$ tighter
  by more than an order of magnitude, ranging from
  $\log_{10}(-f_{R}^0) < -5.81$ to $\log_{10}(-f_{R}^0) < -4.40$
  ($95\%$ confidence level). This sensitivity is due to the different
  shape of the halo mass function, which is degenerate with the
  parameters used to calibrate the scaling relations between SZ
  observables and cluster masses. Any claim of constraints more
  stringent that the weaker limit above, based on cluster number
  counts, appear to be premature and must be supported by a careful
  calibration of the halo mass function and by a robust calibration of
  the mass scaling relations. 
\end{abstract}

\maketitle

Galaxy clusters are the most massive gravitationally bound structures
in the universe ~\citep{allen,review}. The dependence of their
mass- and redshift-dependent number counts is usually described by the halo
\emph{mass function} (MF), $n (M , z)$, the number density of halos in
the mass range $[M, M + dM]$ at redshift $z$. The MF is a sensitive
cosmological probe of the late time universe, and can provide unique
constraints on cosmological parameters and other fundamental physical
quantities, like neutrino masses~\citep{costanzi13,mantz15}. Here we
are interested in constraints on $f(R)$ gravity~\citep{schmidt09a},
which is characterized by a Lagrangian density of the form $R + f(R)$,
where $f$ is a function of the Ricci scalar, $R$. By a conformal
transformation, the fourth order system of field equations can be
reduced to Einstein gravity coupled to a scalar degree of freedom, $f
_{R} = d f / (dR)$, the \emph{scalaron}, of which $f _{R} ^0$
represents the value at current epoch. Another important parameter of
the theory is the present-day Compton wavelength squared of the
scalaron, $B _{0}$, which is proportional to $d ^{2} f / (d R)
^{2}$. Deviations from General Relativity (GR) are quantified by $f$,
and affect gravitational collapse and structure formation, resulting
in a dependence of $n (M, z)$ on $f _{R} ^{0}$. In this \emph{letter}
we discuss viability of $f(R)$ gravity~\citep{pogosian08} by comparing
redshift number counts predictions for galaxy clusters with the
recently released~\citep{Ade:2015fva} all-sky, full-mission,
\emph{Planck} catalogue of Sunyaev-Zel'dovich (SZ) sources (PSZ2), to
constrain $f_{R}^0$.

We describe effects of modified gravity on the cluster MF
following~\citep{Achitouv:2015yha,kopp13}: in these studies $N$-body
simulations are used to fit departure from GR predictions for the
critical density contrast for the collapse of a top-hat spherical
perturbation, $\delta _{\mathrm{c}}$, for models with $f(R) \sim R
^{-n}$, where $n$ is a positive
integer~\cite{hu_sawicki,Starobinsky2007}.  In this context, the good
agreement down to non-linear scales of recent numerical approaches,
which compare theoretical models for the MF ~\cite{lombriser13,
  he13,puchwein13,baldi14,winther15} with the results of different
implementations of $N$-body simulations, motivates the use of an
updated calibration of the MF to improve the robustness of existing
constraints on modified gravity theories~\cite{schmidt09b,cataneo}.


\begin{figure*}
\begin{center}
\includegraphics[width=1.0\textwidth]{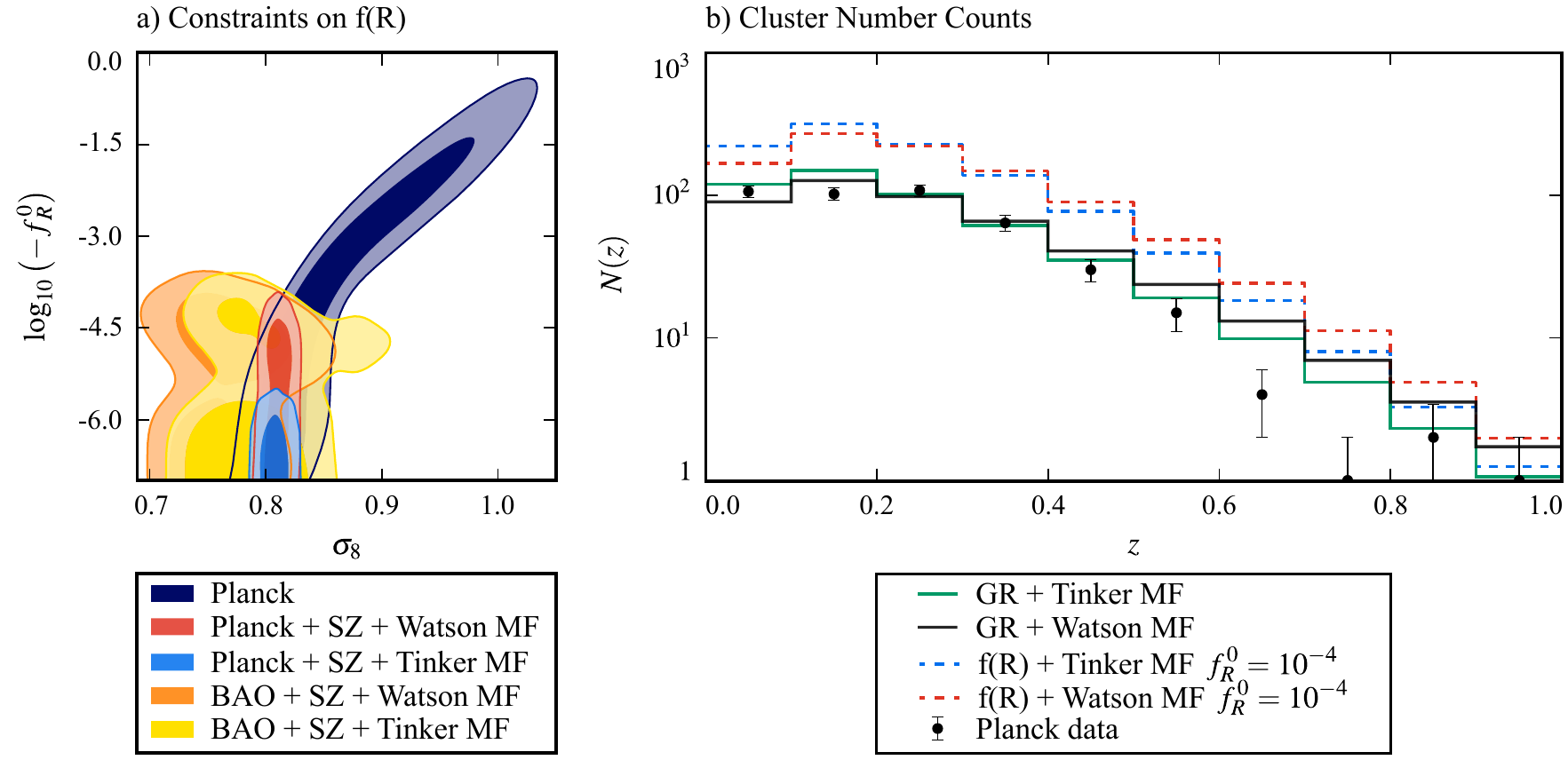}
\caption{{\it Panel a:} the joint marginalized posterior of
  $\log_{10}(-f_{R}^{\,0})$ and $\sigma_8$. Different colors
  correspond to different data set combinations, as shown in
  legend. Constraints that do not include Planck have been obtained by using weak
  priors on n$_{\rm s}$ and  $\Omega_{\rm b}$. 
The darker and lighter shades correspond to the 68\% C.L. and the
95\% C.L. regions, respectively. 
 {\it Panel b:} comparison between the \emph{Planck} measurements and
 the model predictions for the cluster number counts, as a function of redshift. Different colors correspond to different models and different mass functions, as shown in legend. The black data points are samples from the PSZ2 catalogue.
The continuous lines represent the best fit prediction of the \emph{Planck} and \emph{Planck} cluster GR posterior. The dashed lines correspond to the same values of the parameters, but with $\log_{10}(-f_{R}^0)=-4$.
\label{fig_1} }
\end{center}
\end{figure*}

By taking into account Cosmic Microwave Background (CMB) lensing,
constraints from primary CMB temperature anisotropies result in $B
_{0} < 0.1$ at $95\%$ confidence level (C.L.)
\cite{Hu:2014sea}. Adding small scale information from redshift space
distortions and weak lensing~\cite{planck15mg,Hu:2016zrh} further
tightens this constraint to $B _{0} < 0.8 \times 10^{-4}$ ($95\%$
C.L.). Similar results are obtained combining CMB and large scale
structure (e.g., galaxy clustering)
data~\cite{lombriser12,dossett16,bel15,divalentino16}. Here we are
mainly interested in constraints coming from cluster number
counts~\cite{schmidt09b,lombriser12,cataneo}, which have provided upper
limits on $f _{R} ^0$ in the range $[1.3-4.8] \times 10^{-4}$ by
using different data sets and making somewhat different
assumptions. More recently a quite stronger upper limit, $|f _{R} ^0| \lesssim 7 \times
10 ^{-5}$, was obtained from peak statistics in weak
lensing maps~\cite{liu16}.

As for using clusters to derive constraints on cosmological models, a
necessary ingredient is represented by a precise calibration of the
halo MF.  Significant progress in this direction has been made over
the last decade in the context of GR, but only in a few cases
deviations arising in modified gravity theories have been
considered. In general, the MF can be written as~\cite{Bond:1990iw,Press:1973iz}
\begin{equation}
	\frac{d n (M , z)}{d M}
	=
	F (\sigma_M)
	\frac{\rho _{\mathrm{M}}}{M ^{2}}
	\frac{d \log \sigma_M ^{-1}}{d \log M} \,,
\label{eq:M_Fdef}
\end{equation}
where $\rho _{\mathrm{M}}$ is the comoving density of matter, $M$ the
cluster mass, $\sigma_M$ the variance of the linear matter power
spectrum filtered on the mass-scale $M$, and $F$ the
\emph{multiplicity function}. Achitouv \emph{et
  al.}~\cite{Achitouv:2015yha} define a new functional form for $F
(\sigma)$ in $f(R)$ gravity, by fitting results obtained from $N$-body
simulations. This is done by a re-parametrization of $\delta
_{\mathrm{c}}$ that, in contrast with the GR case, becomes scale
dependent and a function of $f_{R}^0$.  

As such, this derivation of the MF for $f(R)$ models should apply 
for halo masses computed at the virial radius. 
In order to calibrate the values of the parameters, Achitouv \emph{et
  al.}~\cite{Achitouv:2015yha} compared their predictions to the MF results
from $f(R)$ $N$-body simulations in the redshift range $z\in [0,1.5]$ and
for scalaron values in the range $-f_{R}^0 \in [10^{-4}$,$10^{-6}]$
\cite{puchwein13}, with halos identified by applying a
\emph{Friends-of-Friends} (FoF) algorithm. 

On the other hand, in the PSZ2 catalogue the cluster masses are given
as $M _{500\mathrm{c}}$, defined as the total mass within a radius, $R
_{500\mathrm{c}}$, chosen in such a way that the mean enclosed density is $500
\rho _{\mathrm{c}}$. In order to adapt this calibration of the $f(R)$
MF to our case, we decided to implement the Achitouv \emph{et al.}~\cite{Achitouv:2015yha} MF as
a correction to the multiplicity function for the GR case, and apply
it to GR  multiplicity functions, computed at $R_{500\mathrm{c}}$, that have been calibrated from large
sets of N-body simulations of standard gravity:
\begin{equation}
	F (\sigma)
	=
	F _{\mathrm{GR}} (\sigma) \,
	\frac{F _{\mathrm{A}} ^{fR} (\sigma)}{F _{\mathrm{A}} ^{\mathrm{GR}} (\sigma)}
	.
\label{eq:f_Rcor}
\end{equation}
Here $F _{\mathrm{A}} ^{fR} (\sigma)$ and $F _{\mathrm{A}}
^{\mathrm{GR}} (\sigma)$ are the multiplicity functions defined in~
Ref.~\cite{Achitouv:2015yha}. As for $F _{\mathrm{GR}} (\sigma)$, the
multiplicity function calibrated on $N$-body simulations in GR, we
implement two alternative definitions: the Tinker \emph{et al.}
MF~\cite{Tinker:2008ff}, and the Watson \emph{et al.}
MF~\cite{Watson:2012mt}. In the following, we will refer to them as
\emph{Tinker} and \emph{Watson}, respectively.
We choose not to test the Achitouv \emph{et al.} MF directly, but in
the form of a correction to another MF, as in eq.~\eqref{eq:f_Rcor},
because its GR limit is markedly different from the Tinker and Watson
results. These two MFs have been widely studied thus allowing us to
compare our results to past literature.

Following this procedure, we are implicitly assuming that the $f(R)$
correction to the MF from Ref. \cite{Achitouv:2015yha} also applies at
$R_{500\mathrm{c}}$. This assumption clearly needs to be verified from an
extensive calibration of the MF at different overdensities from large
$f(R)$ $N$-body simulations

Within the PSZ2 catalogue, we identify a sample of $429$ clusters
with a signal-to-noise ratio $q > 6$. These
clusters have masses in the range $M_{500\mathrm{c}} \in [1,10] \times 10^{14}\,
M_{\odot}$, and redshift $z \in [0, 1]$ and are hereafter denoted as the
\emph{SZ} data set.  The characteristic mass scale of the cluster
sample is a critical element in the number counts analysis. In the
original analysis of the Planck collaboration \cite{Ade:2015fva}, a
calibration of a scaling relation between measured cluster masses and
integrated Compton-$y$ parameter is assumed. To parametrize the uncertain
knowledge in the calibration of cluster masses, a mass bias parameter
$b$ is introduced \cite{Ade:2015fva}, which is defined as the ratio
between the masses calibrated through XMM-Newton X-ray observations
\cite{Arnaud10} and the true cluster masses. In the following, we
assume true cluster masses to be given by the weak lensing results
from the \emph{Weighing the Giants}
project~\cite{vonderLinden:2012kh}. This amounts to assume for the bias
parameter $B _{\mathrm{SZ}} = 1-b$ a Gaussian prior with mean value
$0.688$ and variance $0.072$.  This choice of the bias parameter is
motivated by the fact that it provides a better agreement with primary
\emph{Planck} CMB results. In this sense, this is a conservative
choice since it leaves less freedom for deviations from the standard
$\Lambda$CDM results. By choosing another prior, the tension between
different data sets could result in artificially tighter constraints
on $f_{R}^0$, when combining CMB and cluster number counts
data. Compared to other X-ray selected cluster data sets, like
CCCP~\cite{hoekstra2015} or REFLEX~\cite{reflex1998}, the {\it Planck}
sample is biased towards larger masses and higher redshift, and offers
a unique opportunity to test the MF in a complementary regime. Another
key parameter used in the likelihood analysis is
$\alpha_{\mathrm{SZ}}$, which sets the slope of the scaling relation
between $Y _{500\mathrm{c}}$, the strength of the SZ signal in terms of the
Compton $y$-profile integrated within a sphere of radius $R _{500\mathrm{c}}$,
and $M _{500\mathrm{c}}$.

We also use \emph{Planck} measurements of CMB fluctuations in both
temperature and polarization~\cite{Ade:2015xua, Aghanim:2015xee} in
the multipoles range $\ell \leq 29$. We account for CMB anisotropies
at smaller angular scales by using \texttt{Plik}
likelihood~\cite{Aghanim:2015xee} for CMB measurements of the TT, TE
and EE power spectra. Finally, we include the \emph{Planck} 2015
full-sky lensing potential power spectrum~\cite{Ade:2015xua} in the
multipoles range $40 \leq \ell \leq 400$.
%

We then complement CMB measurements with the \emph{Joint Light-curve
  Analysis} ``JLA'' Supernovae sample, as introduced
in~\cite{Betoule:2014frx}, and with BAO measurements of: the SDSS Main
Galaxy Sample at $z_{\mathrm{eff}} = 0.15$~\cite{Ross:2014qpa}; the
BOSS DR11 ``LOWZ'' sample at $z_{\mathrm{eff}} =
0.32$~\cite{Anderson:2013zyy}; the BOSS DR11 CMASS at $z
_{\mathrm{eff}} = 0.57$~\cite{Anderson:2013zyy}; and the 6dFGS survey
at $z _{\mathrm{eff}} = 0.106$~\cite{Beutler:2011hx}.  We refer to 
this data combination CMB+BAO+JLA as {\it{Planck}}.

We use EFTCAMB and EFTCosmoMC~\cite{Hu:2013twa,Raveri:2014cka},
modifications of the CAMB/CosmoMC
codes~\cite{Lewis:1999bs,Lewis:2002ah}, to compute cosmological
predictions, and compare them with observations.  The EFTCosmoMC code
is modified to account for the $f(R)$ cluster likelihood, which has
been obtained from a suitable modification of the original likelihood
described in \cite{Ade:2015fva}.

\begin{figure*}
\begin{center}
\includegraphics[width=1.0\textwidth]{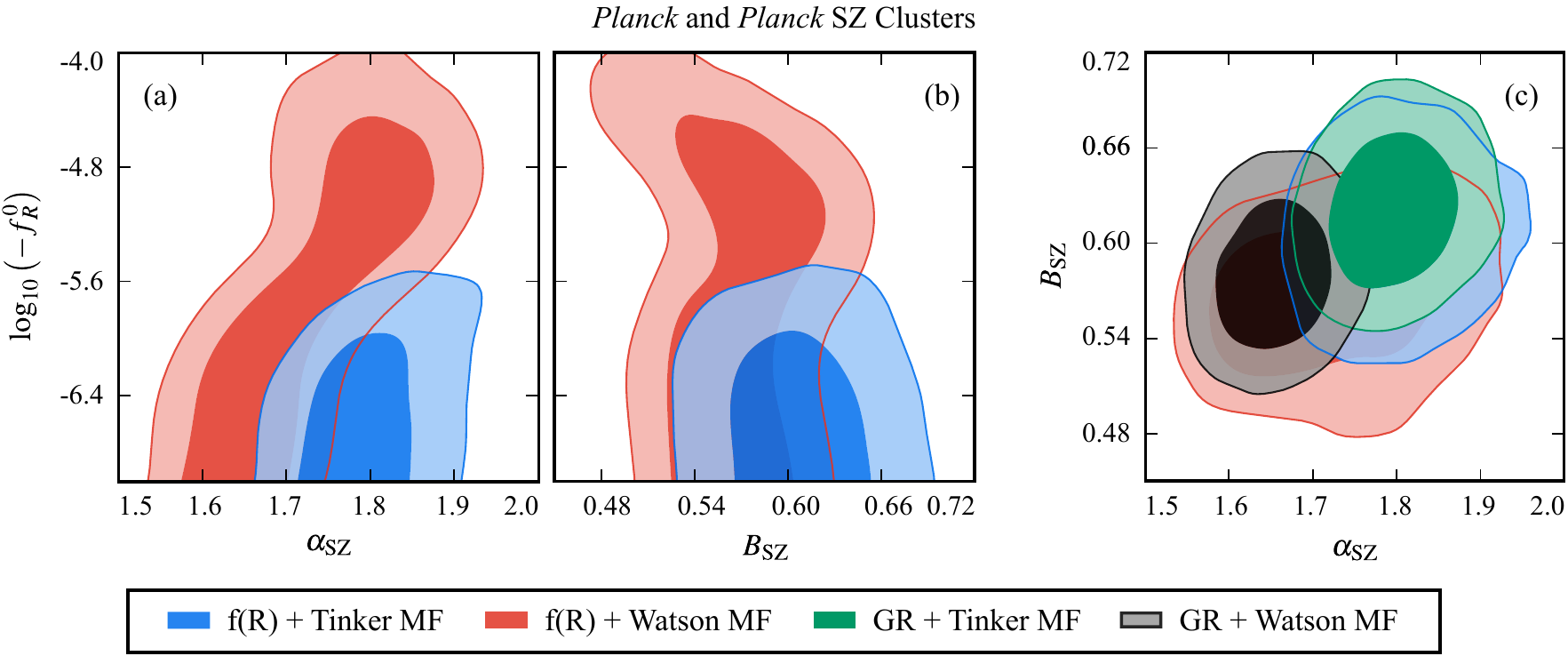}
\caption{The joint marginalized posterior of
  $\log_{10}(-f_{R}^{\,0})$, $\alpha_{\rm SZ}$ and $B_{\rm SZ}$ for
  the \emph{Planck} and \emph{Planck} SZ clusters data sets. Different
  colors correspond to different models, as shown in legend. The
  darker and lighter shades correspond to the 68\%
  C.L. and the 95\% C.L. regions, respectively.
\label{fig_2} 
}
\end{center}
\end{figure*}

\emph{Results}. In Table~\ref{tab:constraints} we show the
marginalized constraints, at $95$\% confidence level, obtained from
the \emph{Planck+SZ} data set, with the $f(R)$ correction applied to,
both, \emph{Tinker} and \emph{Watson} MFs. In the case of the
\emph{Tinker} MF we obtain the tightest constraints on $f(R)$ to date.
In particular we improve the bounds of \cite{cataneo} on
$\log_{10}(-f_{R}^{\,0})$ by one order of magnitude and the ones in
\cite{liu16} by almost an order of magnitude.  In addition these
constraints improve substantially on the bounds coming from large
scale cosmological observations \cite{Hu:2016zrh}, confirming the
leading role of galaxy clusters in constraining modified gravity
theories.

We notice, however, that the upper bound on $\log_{10}(-f_{R}^0)$
strongly depends on the choice of the MF, which can affect
observational constraints by more than one order of magnitude.
\begin{table}[h]
\begin{tabular}{|@{$\;\;$}r@{$\;$}|@{$\;$}c@{$\;$}|@{$\;$}c@{$\;$}|}
\hline
Parameter & {\it Tinker} ($95$ \% C.L.) & {\it Watson} ($95$ \% C.L.)\\
\hline
$\log_{10} (- f_{R}^0)$ & $<-5.81$& $<-4.40$ \\
$\log_{10} B_0$ &  $<-5.60$ & $<-4.06$ \\
$\sigma_8$ & ($0.79$,  $0.83$)& ($0.80$,  $0.83$)\\
$\alpha_{\rm SZ}$ & ($1.68$,  $1.91$)& ($1.57$,  $1.89$)\\
$B_{\rm SZ}$&($0.55$,  $0.67$) & ($0.50$,  $0.63$)\\
\hline
\end{tabular}
\caption{Marginalized constraints at $95$ \%, obtained from the
  \emph{Planck+SZ} data set. Different columns show the two different
  MFs to which the $f(R)$ correction is being applied, see the
  discussion following eq.~ \protect{\eqref{eq:f_Rcor}}. A prior on
  $B_{SZ}$ has been applied as in \citep{Ade:2015fva}. 
\label{tab:constraints}}
\end{table}
This strong dependence is clear also in Fig.~\ref{fig_1}a: the
\emph{Tinker} MF produces the tightest bounds, while the \emph{Watson}
MF is less constraining. Noticeably, SZ cluster measurements break the
degeneracy between $\sigma_8$ and $\log_{10}(-f_{R}^0)$ that
\emph{Planck} CMB measurements clearly display.  Furthermore we
consider a run with SZ clusters without {\it Planck} data, adding the
BAO constraints that we described previously. We also include a prior on
n$_{\rm s}$, n$_{\rm s}=0.9624 \pm 0.014$, taken
from~\cite{planck_xvi} and we adopt Big Bang nucleosynthesis
constraints from~\cite{steigman2008}, $\Omega_{\rm b}=0.022 \pm 0.002$
({\it SZ+BAO} data set).  The results obtained are shown in
Fig.~\ref{fig_1}a, where we report both the {\it Tinker} (in yellow)
and {\it Watson} (in orange) contour plots. We can notice that, at
least for the {\it Tinker} run, the addition of CMB data significantly
improves the constraints on $f_R^{\,0}$ by more than two orders of
magnitude. We also stress that, in the case of {\it SZ+BAO}, we do not
get the strong dependence on the GR calibration of the MF that we
obtain for the {\it SZ+Planck} runs.
In the latter case the constraints obtained for the choice of
\emph{Watson} MF are weaker because the shape of this MF is different
from \emph{Tinker} MF in the range of mass and redshift probed by SZ
\emph{Planck} clusters. More precisely, as shown in~
Fig. \ref{fig_1}b, $N(z)$ falls off at high redshift for the
\emph{Watson} MF more slowly compared to \emph{Tinker} case: when
combined with CMB \emph{Planck} data, in order to fit the tail at high
redshift in GR, a lower $B _{\mathrm{SZ}}$ is required; a lower
$\alpha_{\mathrm{SZ}}$ is instead preferred in order to fit the
low-redshift trend for $N(z)$. 
When we, instead, consider $f(R)$ models for \emph{Watson} MF, there
is a more effective way to change the slope of $N(z)$ with this
parameter (Fig.~\ref{fig_1}) than by using $\alpha_{\mathrm{SZ}}$,
which is now fairly unconstrained and degenerate with $f_{R}^0$. The
same is not true for the \emph{Tinker} case: this degeneracy is not
present, which results in tighter constraints on $f_{R}^0$.

In Figure~\ref{fig_2} we show the contour plots for $f_{R}^0$, and for
the SZ parameters $\alpha_{\mathrm{SZ}}$ and $B _{\mathrm{SZ}}$. This
figure summarizes the interplay between cosmology and astrophysical
parameters of the cluster scaling relations. In the first two panels
we can see the degeneracy between $f_{R}^0$ and the other two parameters:
this is clear in the \emph{Watson} case, but absent in the
\emph{Tinker} one. The wider range of $\alpha_{\rm SZ}$ probed by
Watson MF when compared to Tinker MF in $f(R)$ models is evident and
explains the weaker constraints obtained in the former case.

\emph{Stability of the results}.  To test the dependence of our
results on other effects, we first add the contribution of baryons,
and implement the baryonic correction to the MF discussed
in~\cite{cui14}. In particular, we consider the correction to the MF
obtained when including the effect of feedback from active galactic nuclei 
(AGN) in hydrodynamic simulations. We obtain the constraint
$\log_{10} (-f_R^0) < -5.84$ at $95\%$ C.L. when considering the
\emph{Tinker} MF and the \emph{SZ+Planck} data set. We thus conclude
that the presence of baryons does not have a substantial influence on
our results unlike the larger effects found in Ref.~\cite{costanzi13},
where, however, cluster data probed smaller masses, which are more
affected by feedback effects than those probed by SZ clusters.

We then investigate the dependence from the signal-to-noise ratio of
\emph{Planck} data, by using the most conservative choice $q > 8.5$, that
reduces the sample to 40\% of the original one. In this case
we obtain $\log_{10}(-f_R^0)<-5.54$, at $95\%$ C.L., using
\emph{Tinker} MF, now with this reduced \emph{SZ+Planck} data
set. Again, we can then conclude that our constraints are stable, in
the sense that a change in $q$ affects them much less than a change in
the MF would.

\emph{Discussion}. We compare our results with a recent
work~\cite{cataneo}, where galaxy clusters have been used in order to
get constraints on $f(R)$ gravity theory. In that case the authors got
$\log_{10} (-f_R^0) < -4.79$ by considering the case of {\it Tinker}
MF. In this sense, with the same choice of the MF, our work
improves the constraint by one order of magnitude and gives $\log_{10}
(-f_R^0) < -5.81$. We wish to stress that this result should be
compared with the one in~\cite{cataneo}, since both come from the same
choice of the MF, i.e. {\it Tinker}.  However, the main result of
this {\it letter} goes far beyond the mere exposition of a tighter
constraint. Indeed, we also show that the implementation of a $f(R)$
correction to the MF strongly depends on the calibration of the MF in
GR. In this context we prove that, by keeping the $f(R)$ correction
constant and changing the MF for GR, e.g. by switching from {\it
  Tinker} to {\it Watson}, we obtain a change of more than one order
of magnitude in the $f_R^0$ constraint. In the case of {\it Tinker} we
get $\log_{10} (- f_{R}^0)<-5.81$ (at $95$\% C.L.), while for {\it
  Watson} $\log_{10} (- f_{R}^0)<-4.40$ (at $95$\% C.L.).  As we
already pointed out, this strong dependence on the MF arises from
the degeneracy between $f_R^0$ and the SZ parameters, $\alpha_{\rm
  SZ}$ and $B_{\rm SZ}$. In order to reduce this dependence, it would
be effective to further constrain the clusters mass bias; by
reducing the distribution of this parameter, and thus of $B_{SZ}$, one
would minimize the region of the parameter space in which the
degeneracy occurs. Thus, we expect that a better determination of the
variables describing SZ clusters would directly translate into a more robust
estimation of modified gravity parameters.

In our analysis we also considered stability of the final
results. First of all we implemented the corrections on the MF induced
by considering the effect of baryons and, more specifically, the
effect arising when including star formations and AGN feedback in
hydrodynamic simulations, as described in~\cite{cui14}. In particular
we speculated that the baryonic processes would not depend on the
model of gravity, i.e. on the value of $f_R^0$. In principle, since
these effects strongly influence the shape of the MF and,
consequently, of the cluster number counts, we would expect some
change in the final constraints on the scalaron amplitude. However, as
described before, taking into accounts these effects did not influence
appreciably the result.

We also investigated the effects of the signal-to-noise ratio $q$ for
the identification of the clusters in the \emph{Planck}
catalogue. Setting this threshold to the most conservative one,
$q>8.5$, we obtain $\log_{10}(-f_R^0)<-5.54$, at $95\%$ C. L., which
implies a correction of about $5$\% on the original result for
$\log_{10}(-f_R^0)$. Also in this case, we can then support a
remarkable stability of the result.

In conclusion, we quantitatively investigated the important role that
SZ clusters have in constraining theories of modified gravity once
cluster physics is properly understood and modeled, by using a
state-of-the-art data set and recent results in terms of cluster
MF. While studies in GR are already at an advanced stage, modified
gravity theories can benefit from additional insight on cluster
physics that can be directly translated in tighter constraints on
gravitational physics.  The work presented in this {\it letter} is
thus relevant to present and future cosmological surveys, like Euclid
and CMB-S4, that are expected to deliver unprecedented quality cluster
measurements. A deep understanding of the physics of clusters will
then be the essential to fully exploit the constraining power of these
observations~\cite{mak12,sartoris16}.

\vskip 10pt

\begin{acknowledgments}
%
  We are grateful to Ixandra Achitouv for discussions.  MV is
  supported by the ERC-StG cosmoIGM. MV, SB and MR are supported by
  INFN PD51 - INDARK.  SB and MV are also supported by the PRIN-MIUR
  201278X4FL and by ``Consorzio per la Fisica'' of Trieste. MR
  acknowledges partial support by the Italian Space Agency through the
  ASI contracts Euclid-IC (I/031/10/0).  We thank the Instituut
  Lorentz (Leiden University) and the Osservatorio Astronomico di
  Trieste (OATS) for the allocation of computational resources.
%
\end{acknowledgments}

\bibliographystyle{unsrtnat}
\bibliography{mybib}

\end{document}